\documentclass[12pt,epsf]{article}
\usepackage{amsmath,amssymb}
\usepackage{amsthm,amscd,mathrsfs}
\usepackage{url}
\usepackage{cancel}
\usepackage{ulem}
\usepackage{ascmac}
\usepackage{array}
\usepackage{longtable}
\usepackage{multirow}
\usepackage{arydshln}
\usepackage{bm}

\usepackage{ifpdf}
\ifpdf
  \usepackage[pdftex]{graphicx}
  \usepackage{epstopdf}
\else
  \usepackage[dvips]{graphicx}
\fi

\usepackage{color}
\input{colordvi.tex}

\def\be{\begin{eqnarray}}
\def\ne{\nonumber\end{eqnarray}}
\def\ee{\end{eqnarray}}
\def\nn{\nonumber}

\def\l{\left}
\def\r{\right}
\def\hf{{1\over2}}
\def\i{{i}}
\def\c{\cdot}

\def\sitarel#1#2{\mathrel{\mathop{\kern0pt #1}\limits_{#2}}}
\def\={\buildrel\bigtriangledown\over=}
\def\!={\buildrel!\over=}
\def\?={\buildrel?\over=}
\def\N{{{\cal N}}}
\def\Z{{{\bf Z}}}

\def\1{{{\bf 1}}}
\def\adj{{{\bf adj}}}

\def\matrix#1{\begin{pmatrix}#1\end{pmatrix}}

\def\G{\Gamma}
\def\g{\gamma}

\def\dl{\delta}

\def\la{\lambda}

\def\om{\omega}

\def\s{\sigma}

\def\vev#1{\left\langle{#1}\right\rangle}

\def\tr{{{\rm tr}}}

\newcommand{\drawsquare}[2]{\hbox{%
\rule{#2pt}{#1pt}\hskip-#2pt%
\rule{#1pt}{#2pt}\hskip-#1pt%
\rule[#1pt]{#1pt}{#2pt}}\rule[#1pt]{#2pt}{#2pt}\hskip-#2pt%
\rule{#2pt}{#1pt}}%
\newcommand{\fund}{\raisebox{-.5pt}{\drawsquare{6.5}{0.4}}}
\newcommand{\cpxfund}{\overline{\fund}}
\newcommand{\antisymm}{\raisebox{-3.5pt}{\drawsquare{6.5}{0.4}}\hskip-6.9pt
  \raisebox{3pt}{\drawsquare{6.5}{0.4}}}
\newcommand{\cpxantisymm}{\overline{\antisymm}}
\newcommand{\symm}{\raisebox{-.5pt}{\drawsquare{6.5}{0.4}}\hskip-0.4pt
  \raisebox{-.5pt}{\drawsquare{6.5}{0.4}}}

\begin{document}

\centerline{{\Large\bf Discrete Anomaly Matching}}
\medskip
\centerline{{\Large\bf for the Pouliot Type Dualities}}
\bigskip
\centerline{\sc Teruhiko Kawano}
\bigskip
\centerline{\it Department of Physics, 
Kyoto Prefectural University of Medicine, Kyoto 606-0823, Japan}

\vspace{2cm}
We compute the 't Hooft anomalies of discrete symmetries 
in the Pouliot type dual theories 
and check their anomaly matching conditions. 
The Pouliot type dual theories we will consider in this paper are 
two dual pairs; 
the dual pair of $\N=1$ supersymmetric theories of a $Spin(7)$ gauge theory 
with spinors and a $SU(N_f-4)$ gauge theory with a symmetric tensor, 
fundamentals and singlets, and the other dual pair of 
$\N=1$ supersymmetric theories of a $Spin(10)$ gauge theory with  a spinor 
and vectors and a $SU(N_f-5)$ gauge theory with a symmetric tensor, 
fundamentals and singlets. We will show that the both pairs satisfy 
the discrete anomaly matching conditions.

\vspace{1cm}
\tableofcontents
\thispagestyle{empty}
\clearpage
\addtocounter{page}{-1}

\newpage
\section{Introduction}

In this paper, we will study the 't Hooft anomalies of independent discrete 
symmetries of two dual pairs,  an $\N=1$ supersymmetric $Spin(7)$ 
gauge theory with $N_f$ spinors and its dual \cite{Spin(7)},  and 
an $\N=1$ supersymmetric $Spin(10)$ gauge theory with a spinor and 
$N_f$ vectors and its dual \cite{Spin(10),Kawano}. 
We will show that the two dual pairs satisfy the 't Hooft anomaly 
matching conditions of the independent discrete symmetries. 

The concern that quantum gravitational effects might break global 
symmetries\footnote{See \cite{Ooguri} for recent discussions.}
and the facts that discrete symmetries play important roles 
in phenomenological models have led to the idea of discrete gauge symmetries 
\cite{Krauss-Wilczek}, and further to the anomaly cancellation conditions 
of them \cite{Preskill,Ibanez-Ross,Banks-Dine}\footnote{See \cite{Kobayashi} 
for recent discussions on discrete (non-)Abelian symmetries.}. 
They have been used to derive the anomaly matching conditions of 
discrete symmetries \cite{Csaki-Murayama,Murayama} by extending the discssions 
given by 't Hooft \cite{'t Hooft}. 

The continuous 't Hooft anomaly matching conditions are necessary but severe 
conditions for possible low-energy theories of a strong coupling 
high-energy theory to satisfy. In particular, they yield very strong evidences 
for the Seiberg dualities \cite{Seiberg}\footnote{See \cite{Tachikawa} 
for a recent excellent review.}. The anomaly matching conditions 
for discrete symmetries may give more evidences for the conjectures, 
if a dual pair has independent discrete symmetries. 
In fact, in the papers \cite{Csaki-Murayama,Murayama}, 
for many dual pairs, the discrete anomaly matching have been checked, 
and some of them do not pass the tests. 

Therefore, it is significant to check the discrete anomaly matching 
for unchecked dual pairs. Among those, we will study the above two pairs, 
the $Spin(7)$ theory and its dual, and the $Spin(10)$ theory and its dual. 

In section \ref{Spin(7) dual pair}, we will discuss an independent 
discrete symmetry of the $Spin(7)$ theory and embed the discrete symmetry 
into an anomaly free $U(1)$ symmetry by introducing additional fields 
so as to define the discrete anomalies. 
We will repeat the same discussions for the dual $SU(N_f-4)$ theory. 
In the magnetic theory, the transformation laws of the fields for the discrete 
symmetry is not uniquely determined. However, we will show that 
they all saturate the discrete 't Hooft anomaly matching conditions 
given by the electric $Spin(7)$ theory. 
The dual pair of the $Spin(7)$ theory is obtained by the parent dual pair 
of an $\N=1$ supersymmetric $Spin(8)$ theory with a spinor and $N_f$ vectors 
and its dual \cite{Spin(8)} by higgsing the $Spin(8)$ gauge group 
by the non-zero vacuum expectation value of the spinor. In the $Spin(8)$ theory 
and the dual theory, there is an anomaly free $U(1)$ global symmetry, 
and after the higgsing, it is broken into the discrete symmetry of the 
$Spin(7)$ theory and its dual. The continuous 't Hooft anomalies 
which the $U(1)$ symmetry takes part in become the discrete anomalies, 
and therefore, the fact that 
the parent dual pair satisfies the continuous 't Hooft anomaly matching 
conditions immediately implies that the $Spin(7)$ and its dual 
satisfy the discrete anomaly matching. 
However, there is a subtlety upon defining the discrete anomalies from 
the continuous anomalies because we need to perform a gauge transformation 
as well to gain the discrete symmetry. 
We will elaborate on this issue and show that the dual theories surely 
satisfy the discrete anomaly matching conditions. 

In section \ref{Spin(10) dual pair}, we will proceed to the dual pair of 
the $Spin(10)$ theory, and repeat the same procedure 
as done for the $Spin(7)$ dual pair. We will embed the discrete symmetry 
into an anomaly free $U(1)$ symmetry by adding fields to the theories 
and compute the anomalies. Although we can not uniquely determine 
the transformation laws of the magnetic fields, we will show that 
they all satisfy the discrete anomaly matching conditions, 
as for the $Spin(7)$ dual theories. 
Contrary to the $Spin(7)$ dual theories, there are no known 
parent dual pair, from which the $Spin(10)$ dual pair can be derived. 
Therefore, we haven't found the extended theories with the embedding 
$U(1)$ symmetry of the dual pair, where 
all the continuous 't Hooft anomaly matching conditions 
including those with the embedding $U(1)$ symmetry are satisfied. 
However, in Appendix \ref{appendix}, we will construct extended theories of 
the $Spin(10)$ dual pair, where 
the continuous 't Hooft anomaly matching conditions which 
the embedding $U(1)$ symmetry take part in are satisfied exactly. 
It means that the discrete anomaly matching is automatic after 
the higgsing of the embedding $U(1)$ symmetries on the both sides. 
The rest of the continuous 't Hooft anomaly matching are recovered 
after the decoupling of the additional fields. 

Upon computations of anomalies, we will need to use 
the Dynkin index $T_G(R)$ of a representation $R$ of a group $G$ 
defined by
\be
\tr\l[T_R^a T_R^b\r]=T_G(R)\dl^{ab},
\ne
where $T_R^a$ ($a=1,\cdots,\dim{G}$) are the generators of the group $G$ 
in the representation $R$. 
We often omit the subscript $G$ of the Dynkin index, when it is obvious. 
As for the normalization of $T(R)$, we take $T(\fund)=1$ 
for the fundamental representation $\fund$ of $SU(N)$, 
and $T(\bm{N})=2$ for the vector representation of $Spin(N)$. 
They count the number of zero modes of a single fermion in 
the representation of the group, when the one-instanton background 
is turned on. 
For the adjoint representation of $Spin(N)$, we have $T(\adj)=2(N-2)$. 
The properties of the spin representations frequently depend on the parity of 
$N$ of $Spin(N)$. For the spin reprensentataion ${\bf 2^{n-1}}$ 
of both chiralities of $Spin(2n)$, we have $T({\bf 2^{n-1}})=2^{n-3}$. 
On the other hand, for the spin representation ${\bf 2^{n}}$ 
of $Spin(2n+1)$, $T({\bf 2^{n}})=2^{n-2}$.


\section{The dual pair of the $Spin(7)$ theory with spinors}
\label{Spin(7) dual pair}


We will consider an $\N=1$ supersymmetric $Spin(7)$ gauge theory 
with $N_f$ spinors $Q^i$ ($i=1,\cdots,N_f$) 
with no superpotentials \cite{Spin(7)}. 
Besides the continuous global symmetries $SU(N_f)\times U(1)_R$, 
there is an discrete $\Z_{2N_f}$ symmetry in the theory. 
Under the discrete symmetry transformation, the spinors $Q^i$ 
transform as 
\be
Q^i \quad\to\quad \exp\l({2\pi\i\over2N_f}\r)Q^i. 
\label{Spin(7) ele discrete trf}
\ee
The charge assignments of the spinors $Q^i$ are listed in Table 
\ref{Spin(7) ele}. 
\begin{table}[htb]
\begin{center}
\begin{tabular}{|c|c||c|c|c|}\hline
 & $Spin(7)$ & $SU(N_f)$ & $U(1)_R$ & $\Z_{2N_f}$  \\ \hline
$Q^i$  & $ 8 $ & $N_f$  & $1-5/N_f$ & 1  \\ \hline
\end{tabular}
\end{center}
\caption{The charge assignments of the spinors $Q^i$ in the electric 
$Spin(7)$ theory}
\label{Spin(7) ele}
\end{table}
Performing the discrete transformation (\ref{Spin(7) ele discrete trf}) 
twice, it gives a transformation given by an element of 
the center $\Z_{N_f}$ of the flavor $SU(N_f)$ symmetry. 
It implies that the subgroup $\Z_{N_f} \subset \Z_{2N_f}$ 
can be identified with the center of the flavor group $SU(N_f)$. 
Therefore, we need to take the quotient of $\Z_{2N_f}$ by the subgroup 
$\Z_{N_f}$ to find an independent discrete symmetry.  

When $N_f$ is an odd integer, we have $\Z_{2N_f}\simeq\Z_2\times\Z_{N_f}$, 
where we may identify the cyclic group $\Z_{N_f}$ with the center of the 
flavor group $SU(N_f)$. The matter field $Q^i$ transforms into 
$-Q^i$ under the remaining subgroup $\Z_2$. 
We may identify it with a gauge transformation.
In fact, we may take the gamma matrices for the gauge group $Spin(7)$, 
\begin{equation}
\begin{split}
&\g_1=\s_1\otimes\s_1\otimes\s_1,  \qquad \g_2=\s_2\otimes\s_1\otimes\s_1, 
\cr
&\g_3=\s_3\otimes\s_1\otimes\s_1,  \qquad \g_4=\1_2\otimes\s_2\otimes\s_1,  
\cr
&\g_5=\1_2\otimes\s_3\otimes\s_1,  \qquad \g_6=\1_2\otimes\1_2\otimes\s_2,  
\cr
&\g_7=i\g_1\cdots\g_6=\1_2\otimes\1_2\otimes\s_3,  
\end{split}
\label{Spin(7) gamma}
\end{equation}
so that $\g_{12}=\i\s_3\otimes\1_2\otimes\1_2$, which is one of the generators 
of the $Spin(7)$. Under the gauge transformation generated by $\g_{12}$, 
the matter fields $Q^i$ transform as
\be
Q^i \quad\mapsto\quad \exp\l(\hf\phi\g_{12}\r)Q^i
=\matrix{e^{{\i\over2}\phi}\1_4 & \\ & e^{-{\i\over2}\phi}\1_4}Q^i.
\ne
When $\phi=2\pi$, we see that the gauge transformation yields 
$Q^i \mapsto -Q^i$. Therefore, we have found that the discrete symmetry 
$\Z_{2N_f}$ is a subgroup of $Spin(7)\times{SU}(N_f)$, when $N_f$ 
is an odd integer. It implies that there are no independent discrete symmetries 
for odd $N_f$.

We will compute the discrete 't Hooft anomalies below, whichever 
the $\Z_{2N_f}$ symmetry is independent or not. 
As is explained in \cite{Csaki-Murayama}, 
in order to compute Type II discrete anomalies $\Z_{2N_f}^3$, 
$\Z_{2N_f}U(1)_R^2$, $\Z_{2N_f}^2U(1)_R$, we need to promote the discrete 
$\Z_{2N_f}$ symmetry to a anomaly-free continuous $U(1)$ symmetry. 
To this end, we will extend the theory by introducing a singlet $\Phi$ 
and a spinor $P$ of the gauge group $Spin(7)$. 
Under the promoted $U(1)$ symmetry transformation, the fields transform as 
\be
Q^i ~\to~ \exp\l(\i\om\r)Q^i, 
\quad
\Phi ~\to~ \exp\l(\i2N_f\om\r)\Phi, 
\quad
P ~\to~ \exp\l(-\i{N_f}\om\r)P, 
\ne
and $\Phi$ carries the $U(1)_R$ charge 0,
and $P$ the $U(1)_R$ charge one. 
Then, the $U(1)_R$ charge assignment for the spinors $Q^i$ may be kept 
intact and the $U(1)_R$ symmetry are still anomaly free. 
By inspection, we can verify that the promoted $U(1)$ symmetry is also 
anomaly free. 
We will turn on the superpotential $\Phi{}PP$. 
When we will promote the $U(1)$ symmetry to a gauge symmetry by introducing 
a $U(1)$ gauge superfield and turn on the Fayet-Iliopoulos term 
in the $D$-term potential of the gauged $U(1)$ symmetry, 
we will find the vacuum $\vev{\Phi}\not=0$, where the $U(1)$ symmetry 
is broken into the discrete $Z_{2N_f}$ symmetry. The spinor $P$ 
will gain a mass through the superpotential $\vev{\Phi}PP$ and 
decouple from the rest of the theory in the infrared. 
We can utilize the $U(1)$ gauge superfield to compute the 
Type II discrete anomalies $\Z_{2N_f}^3$, $\Z_{2N_f}U(1)_R^2$, 
$\Z_{2N_f}^2U(1)_R$ by caluculating the 't Hooft anomalies 
$U(1)^3$, $U(1)U(1)_R^2$, $U(1)^2U(1)_R$.  
For the computations, we will also multiply the $U(1)_R$ charges 
by $N_f$ to make them integers. 

\begin{table}[htb]
\begin{center}
\begin{tabular}{|c|c||c|c|c|}\hline
 & $Spin(7)$ & $SU(N_f)$ & $U(1)_R$ & $\Z_{2N_f}$  \\ \hline
$Q^i$  & $ {\bf 8} $ & $\fund$  & $1-5/N_f$ & 1  
\\ \hline\hline
$\Phi$  & $ \1 $ & $\1$  & $0$ & $2N_f$  
\\ \hline
$P$  & $ {\bf 8} $ & $\1$  & $1$ & $-N_f$  
\\ \hline
\end{tabular}
\end{center}
\caption{The charge assignments of the fields in the extended $Spin(7)$ theory}
\label{Spin(7) ele U(1)}
\end{table}

Before proceeding to the computations, we notice that the field with 
the discrete charge $q$ in the representation $R$ of the flavor $SU(N_f)$ 
group and in the representation $R_g$ of the gauge $Spin(7)$ group contributes 
to the discrete anomalies except for the $\Z_{2N_f}SU(N_f)^2$ anomaly 
by a mutiple of $q\dim{R}\dim{R}_g$. We see from Table \ref{Spin(7) ele U(1)} 
that the combination $q\dim{R}\dim{R}_g$ is a multiple of $2N_f$.  
Since we count the discrete anomalies modulo $2N_f$, all the discrete anomalies 
except for the $\Z_{2N_f}SU(N_f)^2$ should be zero modulo $2N_f$. 
In fact, our computations give the discrete 't Hooft anomalies, 
\be
&&\bullet~ \Z_{2N_f}SU(N_f)^2:~8,
\nn\\
&&\bullet~ \Z_{2N_f}(\mbox{gravity})^2:~
1 \times 8N_f + 2N_f \times 1 + (-N_f) \times 8
= 2N_f  
\nn\\
&&\bullet~ \Z_{2N_f}^3:~
\l(1\r)^3\times8N_f
+\l(2N_f\r)^3\times 1 + (-N_f)^3 \times 8
=4\times 2N_f, 
\nn\\
&&\bullet~ \Z_{2N_f}U(1)_R^2:~
1\times\l({-5}\r)^2\times8N_f
+{2N_f}\times\l({-N_f}\r)^2
=(N_f^2+100) \times 2N_f,
\nn\\
&&\bullet~ \Z_{2N_f}^2U(1)_R:~
\l(1\r)^2\times\l({-5}\r)\times8N_f
+\l(2N_f\r)^2\times\l({-N_f}\r)
=-(20+2N_f^2)\times2N_f.
\ne

Note that we can introduce a vector $P'$ of the gauge group 
$Spin(7)$ instead of the spinor $P$ to cancel the $U(1)$ gauge anomaly
and add the term $\Phi{}P'P'$ to the superpotential, 
which becomes the Majorana mass term after the $U(1)$ symmetry breaking 
$\vev{\Phi}\not=0$. Since the vector $P'$ carries the $U(1)_R$ charge 1, 
it contributes only to the discrete 't Hooft anomalies 
$\Z_{2N_f}(\mbox{gravity})^2$ and $\Z_{2N_f}^3$. 
Its contributions increase the anomalies 
$\Z_{2N_f}(\mbox{gravity})^2$ and $\Z_{2N_f}^3$ computed for the spinor $P$ 
by $N_f$ and $N_f^3$, respectively. 

When {$N_f=6$}, the $Spin(7)$ theory is in the confining phase without 
chiral symmetry breaking \cite{Spin(7)}, and the low-energy physics 
is described by the mesons $M^{ij} \sim Q^iQ^j$ and the baryons $B \sim Q^4$ 
with the superpotential
\be
\det{M}-M^{ik}M^{jl}B_{ij}B_{kl}-{\rm Pf}B.
\label{W Nf=6}
\ee
\begin{table}[htb]
\begin{center}
\begin{tabular}{|c|c|c|c|}\hline
 & $SU(N_f=6)$ & $U(1)_R$ & $\Z_{2N_f}$  \\ \hline
$M^{ij}$  & $\symm$  & $2-5/3$ & 2  \\ \hline
$B$  & $\cpxantisymm$  & $4-10/3$ & 4  \\ \hline\hline
$X$ &\1 &0& $-12$
\\ \hline
\end{tabular}
\end{center}
\caption{The low-energy theory extended for $N_f=6$}
\label{LEET}
\end{table}
In order to embed the discrete symmetry into an anomaly free $U(1)$ symmetry, 
we will introduce a singlet $X$, as listed in Table \ref{LEET}, and replace 
the superpotential (\ref{W Nf=6}) by 
\be
X\l(\det{M}-M^{ik}M^{jl}B_{ij}B_{kl}-{\rm Pf}B\r).
\ne
Upon the $U(1)$ symmetry breaking to the discrete symmetry by $\vev{X}\not=0$, 
rescaling the meson $M^{ij}$ and the baryons $B_{ij}$, it is reduced into 
the original low energy theory of $M^{ij}$ and $B_{ij}$. 
Using the fields $M^{ij}$, $B_{ij}$ annd $X$, 
we find that 
the discrete 't Hooft anomalies are given by 
\be
&&\bullet~ \Z_{2N_f}SU(N_f)^2:~
2\times8+{4}\times4=8+2\times12,
\nn\\
&&\bullet~ \Z_{2N_f}(\mbox{gravity})^2:~
{2}\times{6\times7\over2}+{4}\times{6\times5\over2}-12
=6+7\times12,
\nn\\
&&\bullet~ \Z_{2N_f}^3:~
2^3\times{6\times7\over2}
+4^3\times{6\times5\over2}+(-12)^3
=-50 \times 12, 
\qquad
(6^3=18 \times 12),
\nn\\
&&\bullet~ \Z_{2N_f}U(1)_R^2:~
{2}\times\l({-4}\r)^2\times{6\times7\over2}
+{4}\times\l(-{2}\r)^2\times{6\times5\over2}
-12\times(-1)^2
=75 \times 12,
\nn\\
&&\bullet~ \Z_{2N_f}^2U(1)_R:~
\l({2}\r)^2\times\l({-4}\r)\times{6\times7\over2}
+\l({4}\r)^2\times\l({-{2}}\r)\times{6\times5\over2}
+(-12)^2\times(-1)
=-80 \times 12.
\ne
They saturate the discrete 't Hooft anomalies of the $Spin(7)$ theory 
extended by the vector $P'$ instead of the spinor $P$. 
Thus, the low-energy theory of the meson $M^{ij}$ and the baryons $B$ 
passes the discrete 't Hooft anomaly matching tests.

\bigskip

The dual of the $Spin(7)$ theory for $7\leq N_f\leq14$ \cite{Spin(7)} 
is an $\N=1$ supersymmetric $SU(N_f-4)$ gauge theory 
with a $s$ in $\symm$, $N_f$ $\tilde{q}_i$ 
in $\cpxfund$, and singlets $M^{ij}$ with the superpotential 
\be
M^{ij}\tilde{q}_i\c{s}\c\tilde{q}_j+\det{s}. 
\ne
The charge assignments of the fields are listed in Table 
\ref{Spin(7) mag}.

\begin{table}[htb]
\begin{center}
\begin{tabular}{|c|c||c|c|}\hline
 & $SU(N_f-4)$ & $SU(N_f)$ & $U(1)_R$  
 \\ \hline
$s$  & $\symm$ & $1$  & $2/(N_f-4) $    
\\ \hline
$\tilde{q}_i$  & $\cpxfund$ & $\cpxfund$  & $5/N_f-1/(N_f-4)$  
\\ \hline
$M^{ij}$  & $1$ & $\symm$  & $2-10/N_f $  
\\ \hline
\end{tabular}
\end{center}
\caption{The field content of the dual of the $Spin(7)$ theory}
\label{Spin(7) mag}
\end{table}

The correspondence of the 
gauge invariant operators $M^{ij} \sim Q^i\c{Q}^j$, 
$Q^4 \sim \tilde{q}^{N_f-4}$ and the invariance of the superpotential 
determine the transformation laws of the discrete symmetry 
\be
M^{ij} ~\to~ \exp\l(2\pi\i{1 \over N_f}\r)M^{ij}, 
\quad
\tilde{q}_i ~\to~ \exp\l(2\pi\i{2+N_fp \over N_f(N_f-4)}\r)\tilde{q}_i, 
\quad
s ~\to~ \exp\l(-2\pi\i{1+2p \over N_f-4}\r)s, 
\ne
up to an integer $p \in \Z$. The anomaly free condition of the discrete 
symmetry by the gauge interaction is 
\be
{2+N_fp \over N_f(N_f-4)}\times N_f-{1+2p \over N_f-4}\times(N_f-2)
=-(p+1) \in \Z.
\ne
The anomaly free condition for the discrete symmetry is obviously satisfied 
for any integer $p$, 
and it is sufficient for computations of Type I discrete 't Hooft anomalies, 
but, in order to compute Type II discrete anomalies, we need to 
embed the discrete group into an anomaly free $U(1)$ group, as is 
done in the electric theory. To this end, we set $p=-1$ so that 
the gauge anomaly for the discrete symmetry is strictly vanishing\footnote{
This is the simplest solution for embedding of the discrete symmetry into 
a $U(1)$ symmetry. In order to cancel the gauge anomaly, we may introduce 
an appropriate set of massive fields in non-trivial representations of the 
gauge group, as we will see below.}.  
We thus find the transformation laws of the discrete symmetry
\be
\tilde{q}_i \to \exp\l(-2\pi\i{N_f-2 \over N_f(N_f-4)}\r)\tilde{q}_i, 
\qquad
s \to \exp\l({2\pi\i \over N_f-4}\r)s, 
\qquad
M^{ij} \to \exp\l({2\pi\i \over N_f}\r)M^{ij}. 
\ne
We see that the discrete symmetry group is a subgroup of the cyclic group 
$\Z_{N_f(N_f-4)}$. 

Looking at the exponent of the transformed $\tilde{q}_i$, 
\be
-{N_f-2 \over N_f(N_f-4)}=
-{1 \over 2N_f}-{1 \over 2(N_f-4)},
\ne
when we perform the same transformation twice, we see that 
the resulting transformation can be given by an element of the center 
of the flavor group $SU(N_f)$ and an element of the center of the 
gauge group $SU(N_f-4)$. Therefore, it is not an independent discrete symmetry. 
We have seen that this is also the case for the discrete $\Z_{2N_f}$ symmetry 
in the electric theory. In this sence, the independent discrete symmetries 
on the both side of the duality are given by the quotient group $\Z_2$. 
When $N_f$ is an odd integer, we have seen that the quotient group $\Z_2$ 
is given by the center of the gauge group $Spin(7)$ in the electric theory, 
and therefore, no independent discrete symmetry in the electric theory 
are found. In the magnetic theory, when $N_f$ is an odd integer, {\it i.e.}, 
$N_f=2k+1$ ($k \in \Z$), the exponent of the transformed $\tilde{q}_i$ 
may be rewritten into 
\be
-{N_f-2 \over N_f(N_f-4)}=
{k \over N_f}-{k-1 \over N_f-4},
\ne
and also we may rewrite the exponents of the other transformed fields into 
\be
&&\exp\l(2\pi\i{1 \over N_f-4}\r)s=\exp\l(2\pi\i{N_f-3 \over N_f-4}\r)s
=\exp\l(2\pi\i{2(k-1) \over N_f-4}\r)s,
\nn\\[5pt]
&&\exp\l(2\pi\i{1 \over N_f}\r)M^{ij}=\exp\l(2\pi\i{-N_f+1 \over N_f}\r)M^{ij}
=\exp\l(2\pi\i{-2k \over N_f}\r)M^{ij}.
\ne
We find that the transformation laws of the discrete symmetry can be given 
by an element of the center of the flavor group $SU(N_f)$ and an 
element of the center of the gauge group $SU(N_f-4)$. 
Therefore, there is no independent discrete symmetry in the magnetric 
theory as well for odd $N_f$. 

This can be understood in a different way. 
For odd $N_f$, $N_f$ and $N_f-4$ are relatively prime to each other\footnote{
When $N_f$ is odd, it is obivous that $N_f$ is prime to $2^m$, for a 
positive integer $m$. It means that there are two integers $p$, $q$ satisfying 
$N_fp+2^mq=1$. Rewriting it into $N_f(p+q)+(N_f-2^m)(-q)=1$, we see that 
$N_f$ and $N_f-2^m$ are relatively prime to each other. 
}. In the same way, $N_f$ and $N_f-2$ are also relatively prime to each other, 
and shifting $N_f$ by $-2$, we see that 
$N_f-2$ and $N_f-4$ are relatively prime to each other as well. 
Therefore, since $N_f-2$ is prime to $N_f(N_f-4)$, 
the discrete symmetry is a cyclic group $\Z_{N_f(N_f-4)}$. 
Furthermore, as $N_f$ and $N_f-4$ are relatively prime to each other, 
the group $\Z_{N_f(N_f-4)}$ is isomorphic to 
$\Z_{N_f}\times\Z_{N_f-4}$, the product group of the center $\Z_{N_f}$ 
of the flavor group $SU(N_f)$ and the center $\Z_{N_f-4}$ of the gauge 
group $SU(N_f-4)$. 

Although there are no independent discrete symmetries for odd $N_f$, 
we will treat the cases for both odd and even $N_f$ on the same footing. 
We will now promote the discrete $\Z_{N_f(N_f-4)}$ symmetry to 
an anomaly free $U(1)$ symmetry 
by introducing a singlet chiral superfield $X$ 
and replacing the term $\det{s}$ 
in the superpotential by $X\det{s}$. 
We suppose that the singlet $X$ carries no $U(1)_R$ charge 
to leave the $U(1)_R$ charge of the symmetric tensor $s$ unchanged.

The anomaly free condition of the $U(1)$ symmetry by the $SU(N_f-4)$ 
gauge interaction 
and the invariance of the superpotential suggest that 
the fields under the $U(1)$ transformation should transform as 
\be
&&M^{ij} \quad\to\quad e^{(N_f-4)i\om}M^{ij}, 
\qquad 
\tilde{q}_i \quad\to\quad e^{-(N_f-2)i\om}\tilde{q}_i,
\nn\\
&&s \quad\to\quad e^{{N_f}i\om}s,
\qquad 
X \quad\to\quad e^{-{N_f(N_f-4)}i\om}X, 
\ne
with a transformation parameter $\om$. 
Note that the $U(1)$ transformation is distinct from the $U(1)$ transformation 
in the electric theory, as we can verify from the gauge invariant operators 
$M^{ij}\sim Q^iQ^j$, $Q^4 \sim \tilde{q}^{N_f-4}$. However, the reason 
that we embed the discrete symmetry into an anomaly free $U(1)$ symmetry 
is just to define Type II discrete 't Hooft anomalies, but not 
to find the dual theory of the extended electric theory with the extra 
$U(1)$ symmetry. 

Let us introduce a $U(1)$ gauge superfield to 
promote the $U(1)$ symmetry to a gauge symmetry and introduce 
the Fayet-Iliopoulos term in the $D$-term potential of the $U(1)$ gauge 
symmetry so that there exists a vacuum 
$\vev{X}\not=0$, where the $U(1)$ symmetry is broken into 
the original discrete $\Z_{N_f(N_f-4)}$ symmetry. 
As was done for the electric theory, we can make use of the background $U(1)$ 
gauge field to compute Type II discrete anomalies 
$\Z_{2N_f}^3$, $\Z_{2N_f}U(1)_R^2$, $\Z_{2N_f}^2U(1)_R$ 
by computing the 't Hooft anomalies $U(1)^3$, $U(1)U(1)_R^2$, $U(1)^2U(1)_R$.  
We will multiply the $U(1)_R$ charges by $N_f$. However, 
the $U(1)_R$ charges are not all integers even after 
multiplying them by $N_f$, contrary to the electric theory. 

\begin{table}[htb]
\begin{center}
\begin{tabular}{|c|c||c|c|c|}\hline
 & $SU(N_f-4)$ & $SU(N_f)$ & $U(1)_R$  & $\Z_{N_f(N_f-4)}$ 
 \\ \hline
$s$  & $\symm$ & $1$  & $2/(N_f-4) $  & $N_f$
\\ \hline
$\tilde{q}_i$  & $\cpxfund$ & $\cpxfund$  & $5/N_f-1/(N_f-4)$ & $-(N_f-2)$
\\ \hline
$M^{ij}$  & $1$ & $\symm$  & $2-10/N_f $  & $N_f-4$
\\ \hline\hline
$X$ & \1 & \1 & 0 & $-N_f(N_f-4)$
\\ \hline
\end{tabular}
\end{center}
\caption{The field content in the extended magnetic theory}
\label{Spin(7) extended mag}
\end{table}

We compute the 't Hooft anomalies for the magnetic theory, 
\be
&&\bullet~ \Z_{2N_f}SU(N_f)^2:~
-\l({N_f-2}\r) \times (N_f-4)+(N_f-4)\times(N_f+2)=4(N_f-4),
\nn\\[5pt]
&&\bullet~ \Z_{2N_f}(\mbox{gravity})^2:~
N_f\times{(N_f-4)(N_f-3)\over2}
+\l(-(N_f-2)\r)\times N_f \times (N_f-4) 
\nn\\[3pt]
&&\hspace{3.5cm}
+(N_f-4)\times{N_f(N_f+1)\over2}-N_f(N_f-4)
=0, 
\nn\\[5pt]
&&\bullet~ \Z_{2N_f}^3:~
N_f^3\times{(N_f-4)(N_f-3)\over2}
+\l(-(N_f-2)\r)^3\times N_f \times (N_f-4) 
\nn\\[3pt]
&&\hspace{1.8cm}
+(N_f-4)^3\times{N_f(N_f+1)\over2}+\l(-N_f(N_f-4)\r)^3
\nn\\[3pt]
&&\hspace{1.8cm}
=-(N_f^2-1)(N_f-4)^2\times N_f(N_f-4),
\ne
\be
&&\bullet~ \Z_{2N_f}U(1)_R^2:~
N_f\times{(N_f-4)(N_f-3)\over2}\times\l(-N_f+{2N_f\over N_f-4}\r)^2
\nn\\[3pt]
&&\hspace{2.8cm}
+\l(-(N_f-2)\r)\times N_f \times (N_f-4) \times \l(-N_f+5-{N_f \over N_f-4}\r)^2
\nn\\[3pt]
&&\hspace{2.8cm}
+(N_f-4)\times{N_f(N_f+1)\over2}\times\l(N_f-10\r)^2
-N_f(N_f-4)\times(-N_f)^2
\nn\\[3pt]
&&\hspace{2.8cm}
=-4\l(N_f^2-25\r)\times N_f(N_f-4), 
\ne
\be
&&\bullet~ \Z_{2N_f}^2U(1)_R:~
N_f^2\times{(N_f-4)(N_f-3)\over2}\times\l(-N_f+{2N_f\over N_f-4}\r)
\nn\\[3pt]
&&\hspace{2.8cm}
+\l(-(N_f-2)\r)^2\times N_f\times(N_f-4)\times\l(-N_f+5-{N_f \over N_f-4}\r)
\nn\\[3pt]
&&\hspace{2.8cm}
+(N_f-4)^2\times{N_f(N_f+1)\over2}\times\l(N_f-10\r)
+\l(-N_f(N_f-4)\r)^2\times(-N_f)
\nn\\[3pt]
&&\hspace{2.8cm}
=-2\l(N_f^2+5\r)(N_f-4)\times N_f(N_f-4). 
\ne

In order to examine whether the dual theories satisfy the discrete 
't Hooft anomaly matching conditions, we will embed both of 
the discrete $\Z_{2N_f}$ group in the electric theory and 
the discrete $\Z_{N_f(N_f-4)}$ group into a $\Z_{2N_f(N_f-4)}$ group 
by multiplying the discrete charges 
in the electric theory by $N_f-4$ and those in the magnetic theory 
by $2$, respectively. Then, we find that the $\Z_{2N_f(N_f-4)}SU(N_f)^2$ 
anomaly in the electric theory exactly matches the one in the magnetic theory. 
For the other discrete anomalies, we see that the magnetic theory 
saturates the discrete 't Hooft anomaly matching conditions from 
the electric theory, modulo $2N_f(N_f-4)$. 

\begin{table}[htb]
\begin{center}
\begin{tabular}{|c|c||c|c|c|}\hline
 & $SU(N_f-4)$ & $SU(N_f)$ & $U(1)_R$  & $\Z_{2N_f(N_f-4)}$ 
 \\ \hline
$s$  & $\symm$ & $1$  & $2/(N_f-4) $  & $-2(2p+1)N_f$
\\ \hline
$\tilde{q}_i$  & $\cpxfund$ & $\cpxfund$  & $5/N_f-1/(N_f-4)$ & $2(N_fp+2)$
\\ \hline
$M^{ij}$  & $1$ & $\symm$  & $2-10/N_f $  & $2(N_f-4)$
\\ \hline\hline
$X$ & \1 & \1 &0 & $2(2p+1)N_f(N_f-4)$
\\ \hline\hline
$\tilde{X}$ & \1 & \1 & 0 & $-2(p+1)N_f(N_f-4)$
\\ \hline
$F$ & $\fund$ & \1 & $1+1/(N_f-4)$ & $(p+1)N_f(N_f-4)-(1+2p)N_f$
\\ \hline
$\tilde{F}$ & $\cpxfund$ & \1 & $1-1/(N_f-4)$ & $(p+1)N_f(N_f-4)+(1+2p)N_f$
\\ \hline
\end{tabular}
\end{center}
\caption{The field content in the extended magnetic theory for $p\not=-1$}
\label{Spin(7) mag p not=-1}
\end{table}

\bigskip
We have seen the ambiguity of the magnetic discrete symmetry by 
an integer $p$, and we have taken $p=-1$ as the simplest choice 
for an anomaly free $U(1)$ symmetry. We will here consider the case 
$p\not=-1$ and add additional matter fields for an anomaly free $U(1)$ 
symmetry. We read a $U(1)$ symmetry transformation for $p\not=-1$, 
\be
&&M^{ij} ~\to~ e^{i(N_f-4)\om}M^{ij}, 
\qquad
\tilde{q}_i ~\to~ e^{i(2+N_fp)\om}\tilde{q}_i, 
\nn\\
&&s ~\to~ e^{-i(2p+1)N_f\om}s,
\qquad
X ~\to~ e^{i(2p+1)N_f(N_f-1)\om}X, 
\ne
with a transformation parameter $\om$. Since the $U(1)$ symmetry is anomalous, 
we will add additional fields $\tilde{X}$, $F$ and $\tilde{F}$, listed 
in Table \ref{Spin(7) mag p not=-1}, to make it anomaly free. 
We will add the term $\tilde{X}F\tilde{F}$ to the superpotential 
so as to make them massive after the $U(1)$ symmetry breaking. 
Then, the discrete symmetry group is a subgroup of a cyclic group 
$\Z_{2N_f(N_f-4)}$, due to the presence of the extra fundamental pair 
$F$, $\tilde{F}$. The discrete charges of the magnetic fields are listed 
in Table \ref{Spin(7) mag p not=-1}. 

The introduction of a $U(1)$ gauge superfield to promote 
the $U(1)$ symmetry to a $U(1)$ gauge group is the same as for $p=-1$, 
and we will find the vacuum where $\langle{X}\rangle\not=0$, 
$\langle\tilde{X}\rangle\not=0$, breaking the $U(1)$ symmetry to the discrete 
$\Z_{2N_f(N_f-4)}$ symmetry. Multiplying the $U(1)_R$ charges by $N_f$, 
we compute the discrete 't Hooft anomalies. 
We take the $U(1)_R$ charges of $F$ and $\tilde{F}$ to be the values 
in Table \ref{Spin(7) mag p not=-1} in order for the $\Z_{2N_f(N_f-4)}U(1)_R^2$ 
anomaly to be a multiple of $2N_f(N_f-4)$, therefore, saturating the matching 
condition of the anomaly in the electric theory, modulo $2N_f(N_f-4)$, 
as we will see soon. 

To compare these anomalies with those in the electric theory, 
we will embed the electric $\Z_{2N_f}$ group into the $\Z_{2N_f(N_f-4)}$ 
group by multiplying the discrete charges by $N_f-4$, and we find that 
all the discrete 't Hooft anomalies of both electric and magnetic theories 
match modulo $2N_f(N_f-4)$.

\bigskip

In the previous sections, we have promoted the discrete symmetries 
into anomaly free $U(1)$ symmetries by extending both of the dual 
theories to larger theories. However, those extended theories are 
not dual to each other \footnote{For $p\not=0$, we have different 
$U(1)$ symmetries between the dual theories. For $p=0$, although 
we have the same $U(1)$ symmetry, the continuous 't Hooft anomaly 
matching conditions are not satisfied.}. 
But, as we will see below, there is an ideal dual pair of extended theories, 
an $\N=1$ supersymmetric $Spin(8)$ gauge theory with a spinor 
$S$ and $N_f$ vectors $Q^i$ ($i=1,\cdots,N_f$) with no superpotential, and 
its dual theory \cite{Spin(8)}. See Table \ref{Spin(8) ele} for 
the field content of the $Spin(8)$ theory. 

\begin{table}[htb]
\begin{center}
\begin{tabular}{|c|c||c|c|c|}\hline
 & $Spin(8)$ & $SU(N_f)$ & $U(1)$ & $U(1)_R$  \\ \hline
$S$  & $8_s$ & $1$  & $-N_f$ & $0$  \\ \hline
$Q^i$  & $8_v$ & $\fund$  & $1$ & $1-5/N_f$  \\ \hline
\end{tabular}
\end{center}
\caption{The $Spin(8)$ theory with a spinor and $N_f$ vectors}
\label{Spin(8) ele}
\end{table}

Although there are no independent discrete symmetries on the both sides 
of the duality, they have an anomaly free $U(1)$ symmetry. 
After higgsing the $Spin(8)$ to $Spin(7)$ by $\vev{S}\not=0$, 
$N_f$ vectors $Q^i$ of $Spin(8)$ become $N_f$ spinors of $Spin(7)$. 
The dual of the $Spin(8)$ theory is reduced to the dual 
of the $Spin(7)$ theory. Then, the non-zero vacuum expectation value 
$\vev{S}\not=0$ 
also breaks the anomaly free $U(1)$ symmetry to the discrete symmetry 
of the $Spin(7)$ theory. Before the higgsing of the $Spin(8)$ gauge group, 
all the continuous 't Hooft anomaly matching conditions are satisfied by 
the $Spin(8)$ theory and its dual theory. Therefore, 
all the discrete 't Hooft anomaly matching conditions should be satisfied 
by the $Spin(7)$ theory and its dual $SU(N_f-4)$ theory. 
However, it is somewhat less trivial to confirm this, as we will see below. 

The magnetic theory of the $Spin(8)$ theory is an $\N=1$ supersymmetry 
$SU(N_f-4)$ theory with a symmetric tensor $s$, $N_f$ antifundamentals 
$\tilde{q}_i$ ($i=1,\cdots,N_f$), and singlets $M^{ij}$, $X$ 
with the superpotential \cite{Spin(8)}
\be
M^{ij}\tilde{q}_i\c{s}\c\tilde{q}_j+X \det{s}.
\ne
The singlets $M^{ij}$, $X$ correspond to the gauge invariant operators 
of the electric theory, $M^{ij} \sim Q^i\c{Q}^j$, $X \sim S\c{S}$. 
There is also another gauge invariant operator 
$B \sim S^2Q^4 \sim \tilde{q}^{N_f-4}$. 
See Table \ref{Spin(8) mag} for the field content of the magnetic theory. 
\begin{table}[htb]
\begin{center}
\begin{tabular}{|c|c||c|c|c|}\hline
 & $SU(N_f-4)$ & $SU(N_f)$ & $U(1)$ & $U(1)_R$  \\ \hline
$s$  & $\symm$ & $\1$  & $2N_f/(N_f-4)$ & $2/(N_f-4)$  \\ \hline
$\tilde{q}_i$&$\cpxfund$&$\cpxfund$&$-1-N_f/(N_f-4)$&$5/N_f-1/(N_f-4)$  
\\ \hline
$M^{ij}$  & $\1$ & $\symm$  & $2$ & $2-10/N_f$  \\ \hline
$X$  & $\1$ & $\1$  & $-2N_f$ & $0$  \\ \hline
\end{tabular}
\end{center}
\caption{The field content of the dual of the $Spin(8)$ theory}
\label{Spin(8) mag}
\end{table}

The magnetic theory reminds us of the extended dual theory of the 
$Spin(7)$ theory with $p=-1$. In order to go down to the dual theory of 
the $Spin(7)$ theory, we take the vacuum with $\vev{X}\not=0$, which 
breaks the $U(1)$ symmetry to the discrete symmetry, under which 
the fields transform as 
\begin{equation}
\begin{split}
&s ~\to~ \exp\l({2\pi\i\over N_f-4}\r)s, 
\qquad
\tilde{q}_i ~\to~ 
\exp\l[-2\pi\i\l({1\over2N_f}+{1\over2(N_f-4)}\r)\r]\tilde{q}_i, 
\cr
&M^{ij} ~\to~ \exp\l({2\pi\i\over N_f}\r)M^{ij},
\end{split}
\label{Spin(7) p=-1}
\end{equation}
which is identical to the discrete symmetry for $p=-1$. 
Rescaling $s$, $\tilde{q}_i$ and $M^{ij}$, the superpotential 
is reduced into $M^{ij}\tilde{q}_i\c{s}\c\tilde{q}_j+\det{s}$.
Thus, we obtain the dual of the $Spin(7)$ theory. 

On the other hand, in the electric $Spin(8)$ theory, 
the vacuum $\vev{X}\not=0$ corresponds to the vacuum $\vev{S}\not=0$ 
via the correspondence $X \sim S^2$. Then, on the vacuum $\vev{S}\not=0$, 
the gauge group $Spin(8)$ is broken to $Spin(7)$. Since $X$ and $S$ 
transform under the global $U(1)$ symmetry transformation as 
\be
X ~\to~ \exp\l(-4\pi\i{N}_f\om\r)X, 
\qquad
S ~\to~ \exp\l(-2\pi\i{N}_f\om\r)S,
\ne
with a transformation parameter $\om$, the discrete symmetry 
(\ref{Spin(7) p=-1}) corresponds to 
\be
S ~\to~ \exp\l(-\pi\i\r)S, 
\qquad
Q^i ~\to~ \exp\l({2\pi\i\over2N_f}\r)Q^i. 
\label{discrete S}
\ee
In order to leave the vacuum expectation value $\vev{S}\not=0$ invariant, 
we need to perform the gauge transformation 
\be
S ~\to~ -S, 
\qquad 
Q^i ~\to~ Q^i,
\label{gauge S}
\ee
at the same time, which is in a $\Z_2$ subgroup of the center 
of the $Spin(8)$ group. Let us take a closer look at the gauge transformation 
(\ref{gauge S}). To this end, we take the gamma matrices for the gauge group 
$Spin(8)$ 
\be
&&\G_m=\g_m\otimes\s_1,  \qquad (m=1,\cdots,7), 
\qquad
\G_8=\1_8\otimes\s_2
\ne
and the chirality matrix $\G_9=\G_1\G_2\cdots\G_8=\1_8\otimes\s_3$, 
where $\g_m$ ($m=1,\cdots,7$) are the gamma matrices of the $Spin(7)$ group 
in (\ref{Spin(7) gamma}). Let us take the spinor $S$ to be of positive 
chirality $\G_9S=S$. Then, the gauge transformation generated by $\G_{12}$ 
transforms the spinor $S$ and the vectors $Q^i$ into 
\be
S ~\to~ \matrix{e^{{i\over2}\phi}\1_4 & \\ & e^{-{i\over2}\phi}\1_4}S, 
\qquad 
Q^i ~\to~ 
\matrix{\cos\phi & -\sin\phi &  \\ \sin\phi & \cos\phi & \\  & & \1_6 }
Q^i, 
\label{gauge Spin(8)}
\ee
and the $Spin(8)$ gaugino $\la$ into 
\be
\la ~\to~ 
\matrix{\cos\phi & -\sin\phi &  \\ \sin\phi & \cos\phi & \\  & & \1_6 }
\la
\matrix{\cos\phi & \sin\phi &  \\ -\sin\phi & \cos\phi & \\  & & \1_6 }.
\ne
We may define the gauge transformation (\ref{gauge S}) as the limit 
$\phi\to2\pi$, and then it is convenient to decompose the representations 
of the $Spin(8)$ group under $Spin(2) \times Spin(6)$, where 
the $Spin(2)$ group is generated by the $\G_{12}$. The spinor $S$ is decomposed 
into ${\bf 4}_{1/2}\oplus\bar{\bf 4}_{-1/2}$, and the vectors $Q^i$ 
into $\1_{1}\oplus\1_{-1}\oplus{\bf 6}_0$. The gaugino is decomposed into 
$\1_0\oplus{\bf 6}_{1}\oplus{\bf 6}_{-1}\oplus{\bf 15}_0$.
We may regard the charges of the $Spin(2)$ as additional discrete charges 
to accompany with the discrete transformation (\ref{discrete S}) in order to 
leave $\vev{S}\not=0$ invariant, a more refined definition of (\ref{gauge S}). 
The discrete transformation (\ref{discrete S}) is anomaly free,
\be
-\hf \times T({\bf 8}_s) + {1\over 2N_f} \times N_f \times T({\bf 8}_v)=0. 
\ne
Although the naive definition (\ref{gauge S}) of 
the accompanying gauge transformation may be anomalous, 
we can verify that the refined one keeps it anomaly free. 
For example, the spinor $S$ contributes to the gauge anomaly as
\be
\hf \times \l[T_{Spin(6)}({\bf 4})+\l(\hf\r)^2\r]
-\hf \times \l[T_{Spin(6)}(\bar{{\bf 4}})+\l(-\hf\r)^2\r]
=0, 
\ne
with the Dynkin indices $T_{Spin(6)}({\bf 4})$ and $T_{Spin(6)}(\bar{\bf 4})$ 
of the representations ${\bf 4}$ and $\bar{\bf 4}$, 
respectively, of the $Spin(6)$ group, 
where we have used the fact that the generators of the $Spin(6)$ are traceless. 
In the same way, we can show that the contributions from the vectors $Q^i$ 
and the gaugino $\la$ to the gauge anomaly also vanish. 
The global symmetries $SU(N_f)\times{U}(1)_R\times{U}(1)$ of 
the $Spin(8)$ theory are anomaly free, and the generators of the $SU(N_f)$ 
group are traceless. The $Spin(2)$ charges are `traceless' in each of 
the irreducible representations of the $Spin(8)$ group. 
They imply that the contributions from the $Spin(2)$ charges, 
{\it i.e.}, the accompanying gauge transformation, to the discrete 
't Hooft anomalies are canceled to give nothing. For example, 
the $\Z_{2N_f}^3$ anomaly is computed by 
$\l(U(1)+Spin(2)\r)^3$, where $U(1)$ denotes the global $U(1)$ charges, 
which are the same as the discrete charges 
appearing in the discrete transformation (\ref{discrete S}) and $Spin(2)$ 
means the $Spin(2)$ charges of the accompanying gauge transformation. 
Then, $U(1)Spin(2)^2$, $U(1)^2Spin(2)$, and $Spin(2)^3$ 
are zero, and it becomes $U(1)^3$. Thus, we may ignore the $Spin(2)$ 
charges to compute the discrete anomalies, and therefore, 
the discrete 't Hooft anomalies of the $Spin(7)$ theory 
should be given by the continuous 't Hooft anomalies of the $Spin(8)$ theory. 
For the magnetic theories, there is no such subtleties to compute 
the 't Hooft anomalies, and the discrete 't Hooft anomalies of 
the dual of the $Spin(7)$ theory are also given by the continuous 't Hooft 
anomalies of the dual of the $Spin(8)$ theory. 
Since the continuous 't Hooft anomaly matching conditions are satisfied 
\cite{Spin(8)} by the $Spin(8)$ theory and the dual $SU(N_f-4)$ theory, 
the discrete 't Hooft anomaly matching conditions are obviously satisfied 
by the $Spin(7)$ theory and its dual theory. 
In fact, we can verify that the discrete 't Hooft matching conditions 
are satisfied by the direct computations
\be
&&\hspace{-1.2cm}\bullet~ \Z_{2N_f}SU(N_f)^2=U(1)SU(N_f)^2~
\nn\\
&&\hspace{-1cm}\mbox{the electric theory}:~
{1\over2N_f}\times 8={4 \over N_f},
\nn\\
&&\hspace{-1cm}\mbox{the magnetic theory}:~
-\l({1\over2N_f}+{1\over2(N_f-4)}\r)\times(N_f-4)
+{1\over N_f}\times(N_f+2)={4 \over N_f},
\ne
\be
&&\bullet~ \Z_{2N_f}(\mbox{gravity})^2=U(1)(\mbox{gravity})^2~
\nn\\
&&\mbox{the electric theory}:~-\hf \times 8 +{1\over2N_f}\times N_f \times 8
=0,
\nn\\
&&\mbox{the magnetic theory}:~
{1\over N_f-4}\times{(N_f-4)(N_f-3)\over2}
-\l({1\over2N_f}+{1\over2(N_f-4)}\r)\times(N_f-4)\times N_f
\nn\\
&&\hspace{4cm}
+{1\over N_f}\times{N_f(N_f+1)\over2}-1=0,
\ne
\be
&&\bullet~ \Z_{2N_f}^3=U(1)^3~
\nn\\
&&\mbox{the electric theory}:~\l(-\hf\r)^3 \times 8 
+\l({1\over2N_f}\r)^3\times N_f \times 8
={1\over N_f^2}-1,
\nn\\
&&\mbox{the magnetic theory}:~
\l({1\over N_f-4}\r)^3\times{(N_f-4)(N_f-3)\over2}
-\l({1\over2N_f}+{1\over2(N_f-4)}\r)^3\times(N_f-4)\times N_f
\nn\\
&&\hspace{4cm}
+\l({1\over N_f}\r)^3\times{N_f(N_f+1)\over2}-1={1\over N_f^2}-1.
\ne
\be
&&\bullet~ \Z_{2N_f}U(1)_R^2=U(1)U(1)_R^2~
\nn\\
&&\mbox{the electric theory}:~-\hf \times(-1)^2 \times8 
+{1\over2N_f}\times\l(-{5 \over N_f}\r)^2 \times N_f \times 8
={100 \over N_f^2}-4,
\nn\\
&&\mbox{the magnetic theory}:~
{1\over N_f-4}\times\l(-1+{2 \over N_f-4}\r)^2\times{(N_f-4)(N_f-3)\over2}
\nn\\
&&\hspace{4cm}
-\l({1\over2N_f}+{1\over2(N_f-4)}\r)\times
\l(-1+{5 \over N_f}-{1 \over N_f-4}\r)^2\times(N_f-4)\times N_f
\nn\\
&&\hspace{4cm}
+{1\over N_f}\times\l(1-{10 \over N_f}\r)^2\times{N_f(N_f+1)\over2}
-1\times(-1)^2={100 \over N_f^2}-4,
\ne
\be
&&\bullet~ \Z_{2N_f}^2U(1)_R=U(1)^2U(1)_R~
\nn\\
&&\mbox{the electric theory}:~\l(-\hf\r)^2 \times(-1) \times8 
+\l({1\over2N_f}\r)^2\times\l(-{5 \over N_f}\r) \times N_f \times 8
=-{10 \over N_f^2}-2,
\nn\\
&&\mbox{the magnetic theory}:~
\l({1\over N_f-4}\r)^2\times\l(-1+{2 \over N_f-4}\r)\times{(N_f-4)(N_f-3)\over2}
\nn\\
&&\hspace{4cm}
+\l(-{1\over2N_f}-{1\over2(N_f-4)}\r)^2\times
\l(-1+{5 \over N_f}-{1 \over N_f-4}\r)\times(N_f-4)\times N_f
\nn\\
&&\hspace{4cm}
+\l({1\over N_f}\r)^2\times\l(1-{10 \over N_f}\r)\times{N_f(N_f+1)\over2}
+(-1)^2\times(-1)=-{10 \over N_f^2}-2.
\ne

Under the gauge transformation (\ref{gauge Spin(8)}), the gauge invariant 
operator $S^2Q^4$ is left invariant by definition. 
Therefore, under the discrete transformation (\ref{discrete S}), 
it transforms as 
\be
S^2Q^4 ~\to~ \exp\l[2\pi\i\l({4\over2N_f}-1\r)\r]S^2Q^4.
\ne
The contribution from the spinor $S$ to the discrete charge of $S^2Q^4$ 
accounts for the discrepancy of the $U(1)$ symmetries 
between the $Spin(7)$ theory and its dual, upon the promotion 
of the discrete symmetries into anomaly free $U(1)$ symmetries, 
when choosing $p=-1$, as we have seen in the previous sections.  

In summary, we have seen that the discrete 't Hooft anomaly matching between 
the $Spin(7)$ theory and its dual can beautifully be shown by using the 
parent dual pair, the Spin(8) theory and its dual.


\section{The dual pair of the $Spin(10)$ with a spinor and vectors}
\label{Spin(10) dual pair}


We will consider an $\N=1$ supersymmetric $Spin(10)$ gauge theory 
with a single spinor $S$ and $N_f$ vectors $Q^i$ ($i=1,\cdots,N_f$) 
with no superpotentials and its dual theory\cite{Spin(10),Kawano}. 
See Table \ref{Spin(10) ele} for the charge assignments of the matter fields 
in the electric $Spin(10)$ theory. There is an independent $\Z_{2N_f}$ 
symmetry in the theory. 
The discrete $\Z_{2N_f}$ symmetry transformation 
act on the vectors $Q^i$ and the spinor $S$ as 
\be
Q^i \quad\to\quad \exp\l({2\pi\i\over2N_f}\r)Q^i, 
\qquad 
S \quad\to\quad S, 
\ne
which is anomaly free discrete symmetry. 
We will introduce a singlet $\Phi$ and a vector $P$ with the superpotential 
$\Phi{PP}$ in order to promote the discrete symmetry $\Z_{2N_f}$ to an anomaly 
free $U(1)$ symmetry for computations of Type II discrete 't Hooft anomalies 
so that the fields $\Phi$, $P$ cancel the gauge anomaly of the promoted 
$U(1)$ symmetry by the $Spin(10)$ gauge interactions. 
\begin{table}[htb]
\begin{center}
\begin{tabular}{|c|c||c|c|c||c|}\hline
 & $Spin(10)$ & $SU(N_f)$ & $U(1)$ & $U(1)_R$ & $\Z_{2N_f}$ \\ \hline
$S$  & {\bf 16} & $1$  & $-N_f$ & $1$ & $0$ \\ \hline
$Q^i$ & {\bf 10} & $\fund$  & $2$ & $1-8/N_f$ & $1$  \\ \hline\hline
$\Phi$ & {\bf 1} & $1$  & $0$ & $0$ & $2N_f$  \\ \hline
$P$ & {\bf 10} & $1$  & $0$ & $1$ & $-N_f$  \\ \hline
\end{tabular}
\end{center}
\caption{The electric $Spin(10)$ theory with a spinor and $N_f$ vectors}
\label{Spin(10) ele}
\end{table}

In order to make the $U(1)_R$ charges integers, we will multiply them 
by $N_f$. Since the spinor $S$ does not give any contributions to the 
discrete anomalies, we may divide the $U(1)$ charges by two. 
Each of the fields gives a multiple of $q\dim{R}\dim{R_g}$
as its contribution to all the discrete anomalies 
except for $\Z_{2N_f}SU(N_f)^2$, where $q$ is the discrete charge of 
the field, $\dim{R}$ is the dimension of its representation of the 
flavor symmetry group $SU(N_f)$ and $\dim{R_g}$ is the dimension of its 
representation of the gauge group $Spin(10)$. 
As we can see from Table \ref{Spin(10) ele}, the quantity $q\dim{R}\dim{R_g}$ 
for each of all the fields is always a multiple of $2N_f$, 
and therefore, all the discrete anomalies 
except for $\Z_{2N_f}SU(N_f)^2$ are zero modulo $2N_f$. 
In fact, we compute the discrete 't Hooft anomalies; 
\be
&&\bullet~ \Z_{2N_f}SU(N_f)^2:~10,
\nn\\
&&\bullet~ \Z_{2N_f}(\mbox{gravity})^2:~
1 \times 10N_f + 2N_f \times 1 + (-N_f) \times 10
= 2N_f,  
\nn\\
&&\bullet~ \Z_{2N_f}^3:~
1^3\times10N_f
+\l(2N_f\r)^3\times 1 + (-N_f)^3 \times 10
=(-2N_f^2+5)\times 2N_f, 
\nn\\
&&\bullet~ \Z_{2N_f}U(1)_R^2:~
1\times\l({-8}\r)^2\times10N_f
+{2N_f}\times\l({-N_f}\r)^2
=(N_f^2+320) \times 2N_f,
\ne
\be
&&\bullet~ \Z_{2N_f}U(1)^2:~
1\times1^2\times10N_f
=5 \times 2N_f,
\nn\\
&&\bullet~ \Z_{2N_f}U(1)_RU(1):~
1\times\l({-8}\r)\times1\times10N_f
=-40 \times 2N_f,
\nn\\
&&\bullet~ \Z_{2N_f}^2U(1)_R:~
1^2\times\l({-8}\r)\times10N_f
+\l(2N_f\r)^2\times\l({-N_f}\r)
=-(40+2N_f^2)\times2N_f.
\nn\\
&&\bullet~ \Z_{2N_f}^2U(1):~
1^2\times1\times10N_f
=5\times2N_f.
\ne

The magnetic theory exists for $7 \leq N_f \leq 21$, 
and it is an $\N=1$ supersymmetric $SU(N_f-5)$ gauge theory 
with $N_f$ antifundamentals $\tilde{q}_i$, a single fundamental $q$, 
a symmetric tensor $s$ and singlets $M^{ij}$, $Y^i$, 
with the superpotential \cite{Spin(10),Kawano}
\be
M^{ij}\tilde{q}_i\c{s}\c\tilde{q}_j+Y^i\tilde{q}_i\c{q}+\det{s}.
\ne
The charge assignments for the magnetic fields are listed 
in Table \ref{Spin(10) mag}. 
\begin{table}[htb]
\begin{center}
\begin{tabular}{|c|c||c|c|c||c|}\hline
 & $SU(N_f-5)$  & $SU(N_f)$ & $U(1)$ & $U(1)_R$  & $\Z_{2N_f(N_f-5)}$
\\ \hline
$\tilde{q}_i$  & $\cpxfund$  & $\cpxfund$  
& $-2$ & ${8 / N_f}-{1 / (N_f-5)}$ & $-(2N_f-5)$ 
\\ \hline
${q}$  & $\fund$ &   $\1$ 
& $2N_f$ &  $-1+{1 / (N_f-5)}$ & $N_f$
\\ \hline
$s$  & $\symm$ &     $\1$ 
& $0$ &  ${2 / (N_f-5)}$ & $2N_f$
\\ \hline
$M^{ij}$  & ${\bf 1}$ &    $\symm$ & $4$ & $2-{16 / N_f}$ & $2(N_f-5)$
\\ \hline
$Y^{i}$  & ${\bf 1}$ & $\fund$ & $-2(N_f-1)$ & $3-{8 / N_f}$ & $(N_f-5)$
\\ \hline\hline
$X$ & $\1$ & $\1$ & 0 & 0  & $-2N_f(N_f-5)$
\\ \hline
\end{tabular}
\end{center}
\caption{The magnetic theory of the $Spin(10)$ with a spinor and vectors}
\label{Spin(10) mag}
\end{table}
The gauge invariant operators 
$M^{ij}\sim Q^iQ^j$, $Y^i \sim S^2Q^i$, $B \sim S^2Q^5 \sim \tilde{q}^{N_f-5}$ 
are transformed under the discrete symmetry transformation of the electric 
theory as 
\be
&&M^{ij} \to \exp\l({2\pi\i\over N_f}\r)M^{ij},
\qquad
Y^{i} \to \exp\l({2\pi\i\over2N_f}\r)Y^{i},
\qquad
B \to \exp\l(2\pi\i{5\over2N_f}\r)B,
\ne
which, together with the invariance of the superpotential, 
determines the discrete symmetry transformation of the other magnetic fields
\be
&&\tilde{q}_i \to 
\exp\l[2\pi\i\l({1+2p\over2(N_f-5)}-{1\over2N_f}\r)\r]\tilde{q}_i,
\nn\\[5pt]
&&{s} \to \exp\l(-{2\pi\i}{1+2p\over N_f-5}\r){s},
\qquad
{q} \to \exp\l(-{2\pi\i}{1+2p\over2(N_f-5)}\r){q},
\ne
up to an integer $p$. The discrete symmetry transformations with different 
values of $p$ are related to one another by the gauge transformations given 
by elements of the center of the gauge group $SU(N_f-5)$, and therefore, 
there exists only one independent discrete symmetry out of them. 

Since our intension for the promotion of the discrete 
symmetry to an anomaly free $U(1)$ symmetry is just to define Type II 
discrete 't Hooft anomalies in the magnetic theory, we {\it do not} have to 
take the $U(1)$ symmetry to be the promoted $U(1)$ symmetry of the electric 
theory through the correspondence of the gauge invariant operators. 
When we choose $p=-1$ and promote it to the $U(1)$ symmetry 
transformation of the magnetic theory
\be
&&M^{ij} \to \exp\l({2\pi\i\over N_f}\om\r)M^{ij},
\quad
Y^{i} \to \exp\l({2\pi\i\over2N_f}\om\r)Y^{i},
\quad
\tilde{q}_i \to 
\exp\l(-2\pi\i{2N_f-5\over2N_f(N_f-5)}\om\r)\tilde{q}_i,
\nn\\[5pt]
&&{s} \to \exp\l({2\pi\i\over N_f-5}\om\r){s},
\qquad
{q} \to \exp\l({2\pi\i\over2(N_f-5)}\om\r){q},
\ne
with a transformation parameter $\om$, 
we find that the promoted $U(1)$ symmetry is anomaly free, and 
we need to introduce no more fields to cancel the gauge anomaly. 
However, the gauge invariant operator 
$B_{mag} \sim \tilde{q}^{N_f-5}$ transforms under the magnetic $U(1)$ symmetry 
into the transform of $B_{ele} \sim S^2Q^5$ under the electric $U(1)$ symmetry 
multiplied by $e^{-2\pi\i\om}$. 

For the promotion to the anomaly free $U(1)$ symmetry, 
we will also replace  the term $\det{s}$ in the superpotential by 
$X\det{s}$ with a singlet $X$, transforming under the promoted magnetic 
$U(1)$ symmetry as 
\be
X \to \exp\l(-2\pi\i\om\r)X. 
\ne
In order to compute the Type II discrete 't Hooft anomalies, 
we will introduce a $U(1)$ gauge superfield for the promoted magnetic $U(1)$ 
symmetry and will turn on the Fayet-Iliopoulos term in the $D$-term 
potential of the $U(1)$ gauge symmetry. Then, we will find a vacuum with 
$\vev{X}\not=0$, which breaks the $U(1)$ symmetry down back to the original 
discrete symmetry in the magnetic theory. In the infrared, the theory 
is reduced into the original magnetic theory. 

The discrete symmetry group is a subgroup of a cyclic group $\Z_{2N_f(N_f-5)}$. 
However, when we perform the above discrete symmetry transformation twice, 
the resulting transformation can be given by an element of the center 
of the flavor group and an element of the center of the gauge group, 
and therefore, it is not an independent discrete symmetry, anymore. 

As is done for the electric theory, we will multiply the $U(1)_R$ charges 
by $N_f$ and divide the $U(1)$ charges by two. 
Let $\Psi$ be one of the magnetic fields with the discrete charge $q$ 
in the representation $R$ of the flavor symmetry group $SU(N_f)$ and 
in the representation $R_g$ of the gauge symmetry group $SU(N_f-5)$. 
The field $\Psi$ gives its contributions to all the discrete 't Hooft 
anomaly except for $\Z_{2N_f(N_f-5)}SU(N_f)^2$ by a multiple of 
$q\dim{R}\dim{R_g}$. As we can see from Table \ref{Spin(10) mag}, 
the combination $q\dim{R}\dim{R_g}$ is a multiple of $N_f(N_f-5)$, 
and therefore, checking the discrete anomaly matching except for 
$\Z_{2N_f(N_f-5)}SU(N_f)^2$ is to examine whether it is an even or odd 
multiple of $N_f(N_f-5)$ for the anomalies including no $U(1)_R$. 
Although the $U(1)_R$ charges of the magnetic fields are not all 
integers, even after multiplying them by $N_f$, we find that 
the anomalies including the $U(1)_R$ symmetry are also integers 
by the computations, 
\be
&&\bullet~ \Z_{2N_f(N_f-5)}SU(N_f)^2:~
(-2N_f+5)\times(N_f-5)+2(N_f-5)\times(N_f+2)+(N_f-5)
\nn\\
&&\hspace{4.5cm}
=10(N_f-5),
\ne
\be
&&\bullet~ \Z_{2N_f(N_f-5)}(\mbox{gravity})^2:~
(-2N_f+5)\times N_f(N_f-5)+N_f\times(N_f-5)
\nn\\
&&\hspace{4.5cm}
+(2N_f)\times{(N_f-5)(N_f-4)\over2}
+2(N_f-5)\times{N_f(N_f+1)\over2}
\nn\\
&&\hspace{4.5cm}
+(N_f-5)\times N_f+\l(-2N_f(N_f-5)\r)
= 2N_f(N_f-5),  
\nn\\
&&\bullet~ \Z_{2N_f(N_f-5)}^3:~
(-2N_f+5)^3\times N_f(N_f-5)+N_f^3\times(N_f-5)
+(2N_f)^3\times{(N_f-5)(N_f-4)\over2}
\nn\\
&&\hspace{2.7cm}
+\l(2(N_f-5)\r)^3\times{N_f(N_f+1)\over2}
+(N_f-5)^3\times N_f+\l(-2N_f(N_f-5)\r)^3
\nn\\
&&\hspace{2.7cm}
= -(N_f-5)^2(4N_f^2-5)\times2N_f(N_f-5),  
\nn\\[5pt]
&&\bullet~ \Z_{2N_f(N_f-5)}U(1)_R^2:~
(-2N_f+5)\times\l(-N_f+8-{N_f\over N_f-5}\r)^2 \times N_f(N_f-5)
\nn\\
&&\hspace{3.7cm}
+N_f\times\l(-2N_f+{N_f \over N_f-5}\r)^2\times(N_f-5)
\nn\\
&&\hspace{3.7cm}
+(2N_f)\times\l(-N_f+{2N_f \over N_f-5}\r)^2\times{(N_f-5)(N_f-4)\over2}
\nn\\
&&\hspace{3.7cm}
+2(N_f-5)\times\l(N_f-16\r)^2\times{N_f(N_f+1)\over2}
\nn\\
&&\hspace{3.7cm}
+(N_f-5)\times\l(2N_f-8\r)^2 \times N_f+\l(-2N_f(N_f-5)\r)\times(-1)^2
\nn\\
&&\hspace{3.7cm}
= 320\times2N_f(N_f-5),  
\nn\\[5pt]
&&\bullet~ \Z_{2N_f(N_f-5)}U(1)^2:~
(-2N_f+5)\times\l(-1\r)^2 \times N_f(N_f-5)
+N_f\times\l(N_f\r)^2\times(N_f-5)
\nn\\
&&\hspace{3.7cm}
+2(N_f-5)\times2^2\times{N_f(N_f+1)\over2}
+(N_f-5)\times\l(-(N_f-1)\r)^2 \times N_f
\nn\\
&&\hspace{3.7cm}
= \l(N_f^2+5\r)\times2N_f(N_f-5),  
\ne
\be
&&\bullet~ \Z_{2N_f(N_f-5)}U(1)_RU(1):~
(-2N_f+5)\times\l(-N_f+8-{N_f\over N_f-5}\r)\times(-1)\times N_f(N_f-5)
\nn\\
&&\hspace{4.5cm}
+N_f\times\l(-2N_f+{N_f \over N_f-5}\r)\times N_f \times(N_f-5)
\nn\\
&&\hspace{4.5cm}
+2(N_f-5)\times\l(N_f-16\r)\times2\times{N_f(N_f+1)\over2}
\nn\\
&&\hspace{4.5cm}
+(N_f-5)\times\l(2N_f-8\r)^2\times \l(-(N_f-1)\r)\times N_f
\nn\\
&&\hspace{4.5cm}
= -2(N_f^2+20)\times2N_f(N_f-5),  
\nn\\[5pt]
&&\bullet~ \Z_{2N_f(N_f-5)}^2U(1)_R:~
(-2N_f+5)^2\times\l(-N_f+8-{N_f\over N_f-5}\r) \times N_f(N_f-5)
\nn\\
&&\hspace{3.7cm}
+N_f^2\times\l(-2N_f+{N_f \over N_f-5}\r)\times(N_f-5)
\nn\\
&&\hspace{3.7cm}
+(2N_f)^2\times\l(-N_f+{2N_f \over N_f-5}\r)\times{(N_f-5)(N_f-4)\over2}
\nn\\
&&\hspace{3.7cm}
+\l(2(N_f-5)\r)^2\times\l(N_f-16\r)\times{N_f(N_f+1)\over2}
\nn\\
&&\hspace{3.7cm}
+(N_f-5)^2\times\l(2N_f-8\r)\times N_f+\l(-2N_f(N_f-5)\r)^2\times(-1)
\nn\\
&&\hspace{3.7cm}
= -4(N_f-5)(N_f^2+10)\times2N_f(N_f-5),  
\nn\\
&&\bullet~ \Z_{2N_f(N_f-5)}^2U(1):~
(-2N_f+5)^2\times\l(-1\r) \times N_f(N_f-5)
+N_f^2\times\l(N_f\r)\times(N_f-5)
\nn\\
&&\hspace{3.7cm}
+\l(2(N_f-5)\r)^2\times2\times{N_f(N_f+1)\over2}
+(N_f-5)^2\times\l(-(N_f-1)\r) \times N_f
\nn\\
&&\hspace{3.7cm}
= 5\l(N_f-5\r)\times2N_f(N_f-5). 
\ne

In order to check the discrete 't Hooft matching conditions between 
the dual theories, we will embed the $\Z_{2N_f}$ symmetry group 
of the electric theory into the $\Z_{2N_f(N_f-5)}$ group 
by multiplying the discrete charges of the electric fields by 
$N_f-5$. Then, we can see that the $\Z_{2N_f(N_f-5)}SU(N_f)^2$ anomaly 
in the electric theory is in agreement with the one in the magnetic theory. 
Recalling that the other discrete anomalies in the electric theory 
are zero modulo $2N_f(N_f-5)$, we find that the 't Hooft anomaly matching 
conditions of them are also satisfied by the magnetic theory.

Incidentally, let us consider the discrete 't Hooft anomalies for the other 
cases with $p\not=-1$. When $n \equiv p+1 \not=0$, 
the invariance of the term $X\det{s}$ in the superpotential requires 
the discrete charge of $X$ to be $-2N_f(N_f-5)(1-2n)$, 
as in Table \ref{Spin(10) mag p not=-1}. Then, the promoted $U(1)$ symmetry 
is anomalous by the $SU(N_f-5)$ gauge interactions, and therefore, we need to 
introduce more matter fields listed in Table \ref{Spin(10) mag p not=-1} 
to cancel the gauge anomaly in order to promote the discrete $\Z_{2N_f(N_f-5)}$ 
symmetry to an anomaly $U(1)$ symmetry. 
We will also add the term $\tilde{X}F\tilde{F}$ to the superpotential so that 
the extra matter fields $\tilde{X}$, $F$, $\tilde{F}$ decouple in the vacuum 
$\langle{\tilde{X}}\rangle\not=0$ at the low energies. 
The discrete charges of $F$ and 
$\tilde{F}$ are determined by cancellation of the gauge anomaly of the promoted 
$U(1)$ symmetry and by the requirement that performing 
the discrete symmetry transformation twice gives the gauge transformation 
of the element of the center of the gauge group $SU(N_f-5)$. 
We have chosen the $U(1)_R$ charges of $F$ and $\tilde{F}$ so as to 
saturate the 't Hooft anomaly matching condition 
for $\Z_{2N_f(N_f-5)}U(1)_R^2$. 

\begin{table}[htb]
\begin{center}
\begin{tabular}{|c|c||c|c|c||c|}\hline
 & $SU(N_f-5)$  & $SU(N_f)$ & $U(1)$ & $U(1)_R$  & $\Z_{2N_f(N_f-5)}$
\\ \hline
$\tilde{X}$  & $\1$  & $\1$  
& $0$ & $0$ & $-2N_f(N_f-5)n$ 
\\ \hline
$F$  & $\fund$ &   $\1$ 
& $0$ &  $1+{1 / (N_f-5)}$ & $nN_f(N_f-5)+N_f$
\\ \hline
$\tilde{F}$  & $\cpxfund$ &     $\1$ 
& $0$ &  $1-{1 / (N_f-5)}$ & $nN_f(N_f-5)-N_f$
\\ \hline \hline
${X}$  & $\1$  & $\1$  
& $0$ & $0$ & $-2N_f(N_f-5)(1-2n)$ 
\\ \hline
\end{tabular}
\end{center}
\caption{The discrete charge of $X$ and 
the additional fields $\tilde{X}$, $F$, $\tilde{F}$ 
for $p\not=-1$ in the magnetic dual of the $Spin(10)$ theory, where $n=p+1$.}
\label{Spin(10) mag p not=-1}
\end{table}

The remaining procedure we have to carry out for the computation of the 
discrete anomalies is almost the same as what was done for $p=-1$, 
except that we take the vacuum such that $\vev{X}\not=0$, 
$\langle\tilde{X}\rangle\not=0$.

The $\Z_{2N_f(N_f-5)}SU(N_f)^2$ anomaly for $p\not=-1$ is computed to give 
\be
\bullet~ \Z_{2N_f(N_f-5)}SU(N_f)^2:~ 10(N_f-5)+2N_f(N_f-5)n,
\ne
which is equal modulo $2N_f(N_f-5)$ to the one for the case $p=-1$ ($n=0$). 
For the rest of the discrete anomalies, multiplying the $U(1)_R$ charges 
by $N_f$ and dividing the $U(1)$ charges by two, we find that 
they are all zero modulo $2N_f(N_f-5)$, satisfying all the 't Hooft 
anomaly matching conditions.

\section{Discussions}

We have studied the discrete anomaly matching 
of the two dual pairs. One of them is the $Spin(7)$ gauge theory 
with spinors and the $SU(N_f-4)$ gauge theory with a symmetric tensor, 
fundamentals and singlets \cite{Spin(7)}. 
The other is the $Spin(10)$ gauge theory 
with a spinor and vectors and the $SU(N_f-5)$ gauge theory with a symmetric 
tensor, fundamentals and singlets \cite{Spin(10), Kawano}. 
We have shown that both of the dual pairs satisfy the discrete anomaly 
matching conditions. 

For the dual pair of the $Spin(7)$ theory, we have done this in two ways. 
In one way, we have embedded the discrete symmetries into 
an anomaly free $U(1)$ symmetries by additional fields, 
which decouple after the $U(1)$ symmetry breaking into the discrete symmetries, 
on the both sides of the duality. The extended theories are not dual to each 
other, and we have to compute the continuous 't Hooft anomalies 
in order to ensure the anomaly matching conditions. 
In the other way, we take another dual pair \cite{Spin(8)} of 
the $Spin(8)$ gauge theory with a spinor and vectors and 
the $SU(N_f-4)$ gauge theory with a symmetric tensor, anti-fundamentals 
and singlets, which is reduced to the $Spin(7)$ dual pair by 
higgsing the $Spin(8)$ gauge group to $Spin(7)$. 
The $Spin(8)$ dual pair has an anomaly free $U(1)$ symmetry, 
which is broken to the discrete symmetries of the $Spin(7)$ dual pair 
upon the higgsing of $Spin(8)$. Since the continuous 't Hooft anomaly matching 
is satisfied by the $Spin(8)$ dual pair and since 
the discrete anomalies of the $Spin(7)$ dual pair 
are given by the continuous anomalies of the $Spin(8)$ dual pair, 
we have seen that the discrete anomaly matching conditions are 
obviously satisfied by the $Spin(7)$ dual pair, modulo subtleties 
with the gauge transformation involved in the actual computations. 

For the $Spin(10)$ dual pair, we have no known parent dual pair 
and we have computed the discrete anomalies by embedding the discrete 
symmetries on the both sides of the duality into anomaly free $U(1)$ 
symmetries. In Appendix \ref{appendix}, we have constructed the extended 
theories with the embedding $U(1)$ symmetries, which satisfy the 
continuous 't Hooft anomaly matching conditions for 
the anomalies with the $U(1)$ symmetries. Although the $U(1)$ 
symmetries of the both sides of the duality are different from 
each other, the continuous anomalies with the $U(1)$ symmetries 
become the discrete anomalies after the higgsing of the $U(1)$ symmetries, 
and we easily see that the discrete anomaly matching is achieved.

Another dual pair of an $\N=1$ supersymmetric $G_2$ gauge theory 
with fundamentals in the representation ${\bf 7}$
and an $\N=1$ supersymmetric $SU(N_f-3)$ gauge theory 
with a symmetric tensor, fundamentals and singlets 
is reduced from the $Spin(7)$ dual pair \cite{Spin(7)}\footnote{See 
also \cite{G2}, for the confining phases of the $G_2$ gauge theory.} 
by higgsing the $Spin(7)$ gauge group. 
Therefore, it should be straightforward to check the discrete anomaly 
matching in a similar way to what we have done for 
the $Spin(8)$ dual pair. 
This is also the case for 
other Pouliot type dualities \cite{Cho} reduced from the 
$Spin(10)$ dual pair.

There are no known dual pair, from which the $Spin(10)$ dual pair 
is derived. Although we wish that the extended theories 
in Appendix \ref{appendix} would give an insight into the discovery 
of such a parent dual pair, we guess that the gauge group of the electric 
$Spin(10)$ theory should be larger than the $Spin(10)$ group 
\footnote{See \cite{SO} for the confining phases of 
$\N=1$ supersymmetric $Spin(11)$ and $Spin(12)$ gauge theories 
with a spinor and vectors.}, 
so that the higgsing in the electric theory should correspond to 
the decoupling of massive states in the magnetic theory. 

Finally, we may extend the studies in this paper to a dual pair of an $\N=1$ 
supersymmetric $Spin(10)$ gauge theory with more than one spinor 
and vectors, and the dual theory \cite{Strassler}. 
However, we will leave this subject for future investigation. 

\bigskip\bigskip\bigskip
\centerline{{\bf Acknowledgement}}
The author thanks Yukawa Institute for Theoretical Physics, Kyoto University 
for hospitality during the course of this project. 

\newpage
\appendix
\noindent{\LARGE {\bf Appendix}}

\section{The continuous 't Hooft anomaly matching of the extended theories 
of the $Spin(10)$ dual pair}
\label{appendix}

Contrary to the $Spin(7)$ dual pair, 
there is no known parent dual pair, which is reduced into 
the $Spin(10)$ theory and the dual $SU(N_f-5)$ theory by 
higgsing or decoupling of massive states. 
Therefore, there are no reasons to expect that the extended theories 
of the $Spin(10)$ theory and its dual satisfy all the continuous 
't Hooft anomaly matching conditions. 
The extended theories have the $U(1)$ symmetries, 
into which the discrete symmetries are embedded. 
We will denote the $U(1)$ symmetries as $U(1)_X$, 
although the $U(1)_X$ symmetry of the electric theory is not the same 
as the one of the magnetic theory, for $p\not=0$. 

For $p=-1$, we will compute the continuous 't Hooft anomalies 
of the both extended theories, and will find that there are discrepancies 
in the anomalies with the $U(1)_X$ charges between the electric theory 
and the magnetic theory. However, we will show that we can eliminate 
the discrepancies by adding a set of fields to the electric theory, 
as shown in Table \ref{U(1)_X additional fields ele} 
and to the magnetic theory as listed in Table 
\ref{U(1)_X additional fields mag}. 
After higgsing the $U(1)_X$ symmetry down to the discrete symmetry, 
the additional fields gain masses and decouple in the infrared. 
Then, the anomalies with the $U(1)_X$ charges become the discrete 't Hooft 
anomalies of the original dual theories, and we can ensure that 
they satisfy the discrete 't Hooft anomaly matching conditions. 

The anomalies without the $U(1)_X$ charges 
are the same as those of the original dual theories, except for the 
$U(1)_R^3$ and $U(1)_R(\mbox{gravity})^2$ anomalies. 
Therefore, they satisfy the 't Hooft anomaly matching conditions. 
For the $U(1)_R^3$ and $U(1)_R(\mbox{gravity})^2$ anomalies, 
there are additional contributions from the Higgs fields $\Phi$ 
in the electric theory and $X$ in the magnetic theory, respectively. 
However, their contributions are the same, and 
the $U(1)_R^3$ and $U(1)_R(\mbox{gravity})^2$ anomalies also satisfy 
the anomaly matching conditions between the electric theory and the magnetic 
theory. The remaining anomalies are those which the $U(1)_X$ charges 
take part in, and our results of the computations of them are the following:

\bigskip
\noindent
$\bullet$ $U(1)_XSU(N_f)^2$:~
\be
&&\mbox{the electric side}:~{5 \over N_f},
\nn\\
&&\mbox{the magnetic side}:~
\l(-{1\over2N_f}-{1 \over 2(N_f-5)}\r)(N_f-5)+
{2\over2N_f}(N_f+2)+{1\over2N_f}={5 \over N_f}.
\ne

\noindent
$\bullet$ $U(1)_X(\mbox{gravity})^2$:~
\be
&&\mbox{the electric side}:~
{1\over2N_f}\times10N_f+1-\hf\times10=1,
\nn\\[5pt]
&&\mbox{the magnetic side}:~
\l(-{1\over2N_f}-{1 \over 2(N_f-5)}\r)N_f(N_f-5)
+\l({1 \over 2(N_f-5)}\r)(N_f-5)
\nn\\[5pt]
&&\hspace{3.8cm}
+\l({2 \over 2(N_f-5)}\r){(N_f-5)(N_f-4)\over2}
+{2\over2N_f}{N_f(N_f+1)\over2}+{1\over2N_f}N_f-1
\nn\\[5pt]
&&\hspace{3.8cm}
=1.
\ne

\noindent
$\bullet$ $U(1)_X^3$:~
\be
&&\mbox{the electric side}:~
\l({1\over2N_f}\r)^3\times10N_f+1^3+\l(-\hf\r)^3\times10
={5\over4N_f^2}-{1\over4},
\nn\\
&&\mbox{the magnetic side}:~
\l(-{1\over2N_f}-{1 \over 2(N_f-5)}\r)^3N_f(N_f-5)
+\l({1 \over 2(N_f-5)}\r)^3(N_f-5)
\nn\\[5pt]
&&\hspace{3.8cm}
+\l({2 \over 2(N_f-5)}\r)^3{(N_f-5)(N_f-4)\over2}
+\l({2\over2N_f}\r)^3{N_f(N_f+1)\over2}
\nn\\[5pt]
&&\hspace{3.8cm}
+\l({1\over2N_f}\r)^3N_f-1
={5\over4N_f^2}-1.
\ne

\noindent
$\bullet$ $U(1)_XU(1)_R^2$:~
\be
&&\mbox{the electric side}:~{1 \over 2N_f}\times\l(-{8 \over N_f}\r)^2 
\times10N_f+1\times(-1)^2=320{1 \over N_f^2}+1,
\nn\\
&&\mbox{the magnetic side}:~
\l(-{1\over2N_f}-{1 \over 2(N_f-5)}\r)
\l(-1+{8 \over N_f}-{1\over N_f-5}\r)^2N_f(N_f-5)
\nn\\[5pt]
&&\hspace{3.8cm}
+\l({1 \over 2(N_f-5)}\r)\l(-2+{1\over N_f-5}\r)^2(N_f-5)
\nn\\[5pt]
&&\hspace{3.8cm}
+\l({2 \over 2(N_f-5)}\r)\l(-1+{2\over N_f-5}\r)^2{(N_f-5)(N_f-4)\over2}
\nn\\[5pt]
&&\hspace{3.8cm}
+{2\over2N_f}\l(1-{16 \over N_f}\r)^2{N_f(N_f+1)\over2}
+{1\over2N_f}(2-{8 \over N_f})^2N_f-1\times(-1)^2
\nn\\[5pt]
&&\hspace{3.8cm}
=320{1 \over N_f^2}.
\ne

\noindent
$\bullet$ $U(1)_XU(1)^2$:~
\be
&&\mbox{the electric side}:~{1\over2N_f}\times2^2\times10N_f
=20,
\nn\\
&&\mbox{the magnetic side}:~
\l(-{1\over2N_f}-{1 \over 2(N_f-5)}\r)\l(-2\r)^2N_f(N_f-5)
\nn\\[5pt]
&&\hspace{3.8cm}
+\l({1 \over 2(N_f-5)}\r)\l(2N_f\r)^2(N_f-5)
+{2\over2N_f}\times4^2\times{N_f(N_f+1)\over2}
\nn\\[5pt]
&&\hspace{3.8cm}
+{1\over2N_f}\Big(-2(N_f-1)\Big)^2N_f
=20+4N_f^2.
\ne

\noindent
$\bullet$ $U(1)_XU(1)_RU(1)$:~
\be
&&\mbox{the electric side}:~{1 \over 2N_f}\times2\times\l(-{8 \over N_f}\r) 
\times10N_f=-{80 \over N_f},
\nn\\
&&\mbox{the magnetic side}:~
\l(-{1\over2N_f}-{1 \over 2(N_f-5)}\r)(-2)\l(-1+{8 \over N_f}-{1\over N_f-5}\r)
N_f(N_f-5)
\nn\\[5pt]
&&\hspace{3.8cm}
+\l({1 \over 2(N_f-5)}\r)(2N_f)\l(-2+{1\over N_f-5}\r)(N_f-5)
\nn\\[5pt]
&&\hspace{3.8cm}
+{2\over2N_f}4\l(1-{16 \over N_f}\r){N_f(N_f+1)\over2}
+{1\over2N_f}\big(-2(N_f-1)\big)\l(2-{8 \over N_f}\r)N_f
\nn\\[5pt]
&&\hspace{3.8cm}
=-{80 \over N_f}-4N_f.
\ne

\noindent
$\bullet$ $U(1)_X^2U(1)_R$:~
\be
&&\mbox{the electric side}:~\l({1 \over 2N_f}\r)^2\times\l(-{8 \over N_f}\r) 
\times10N_f=-{20 \over N_f^2},
\nn\\
&&\mbox{the magnetic side}:~
\l(-{1\over2N_f}-{1 \over 2(N_f-5)}\r)^2\l(-1+{8 \over N_f}-{1\over N_f-5}\r)
N_f(N_f-5)
\nn\\[5pt]
&&\hspace{3.8cm}
+\l({1 \over 2(N_f-5)}\r)^2\l(-2+{1\over N_f-5}\r)(N_f-5)
\nn\\[5pt]
&&\hspace{3.8cm}
+\l({2 \over 2(N_f-5)}\r)^2\l(-1+{2\over N_f-5}\r){(N_f-5)(N_f-4)\over2}
\nn\\[5pt]
&&\hspace{3.8cm}
+\l({2\over2N_f}\r)^2\l(1-{16 \over N_f}\r){N_f(N_f+1)\over2}
+\l({1\over2N_f}\r)^2\l(2-{8 \over N_f}\r)N_f
\nn\\[5pt]
&&\hspace{3.8cm}
=-{20 \over N_f^2}-1.
\ne

\noindent
$\bullet$ $U(1)_X^2U(1)$:~
\be
&&\mbox{the electric side}:~
\l({1\over2N_f}\r)^2\times2\times10N_f+1^2\times(-1)={5 \over N_f}-1,
\nn\\
&&\mbox{the magnetic side}:~
\l(-{1\over2N_f}-{1 \over 2(N_f-5)}\r)^2\l(-2\r)N_f(N_f-5)
\nn\\[5pt]
&&\hspace{3.8cm}
+\l({1 \over 2(N_f-5)}\r)^2\l(2N_f\r)(N_f-5)
+\l({2\over2N_f}\r)^24{N_f(N_f+1)\over2}
\nn\\[5pt]
&&\hspace{3.8cm}
+\l({1\over2N_f}\r)^2\Big(-2(N_f-1)\Big)N_f+(-1)^2\times(-1)
={5 \over N_f}-1.
\ne

\begin{table}[htb]
\begin{center}
\begin{tabular}{|c|c||c|c|c||c|}\hline
 & $Spin(10)$ & $SU(N_f)$ & $U(1)$ & $U(1)_R$ & $U(1)_X$ \\ \hline
$S$  & {\bf 16} & $\1$  & $-N_f$ & $1$ & $0$ \\ \hline
$Q^i$ & {\bf 10} & $\fund$  & $2$ & $1-8/N_f$ & $1/(2N_f)$  
\\ \hline
$\Phi$ & {\bf 1} & $\1$  & $0$ & $0$ & $1$  \\ \hline
$P$ & {\bf 10} & $\1$  & $0$ & $1$ & $-1/2$  
\\ \hline\hline
$\tilde{\Phi}$ & \1 & \1 & 0 & 0 & $-1$
\\ \hline
${\Psi}$ & \1 & \1 & 0 & 1 & 1/2
\\ \hline
$\tilde{\Psi}$ & \1 & \1 & 0 & 1 & 1/2
\\ \hline\hline
${\Xi}$ & \1 & \1 & 0 & 0 & $-2$
\\ \hline
${G}$ & \1 & \1 & $-N_f$ & 2 & 2
\\ \hline
${H}$ & \1 & \1 & $N_f$ & 0 & 0
\\ \hline
$\tilde{\Xi}$ & \1 & \1 & 0 & 0 & $-2$
\\ \hline
$\tilde{G}$ & \1 & \1 & $N_f$ & 0 & 2
\\ \hline
$\tilde{H}$ & \1 & \1 & $-N_f$ & 2 & 0
\\ \hline
\end{tabular}
\end{center}
\caption{The original fields and 
the set of the fields added to the extended $Spin(10)$ theory 
to fill the gaps in the continuous 't Hooft anomalies}
\label{U(1)_X additional fields ele}
\end{table}

We see that there are discrepancies in the anomalies except for 
$U(1)_XSU(N_f)$, $U(1)_X(\mbox{gravity})^2$ and $U(1)_X^2U(1)$, 
between both the sides. We will add the singlet fields $\tilde{\Phi}$, 
$\Psi$ and $\tilde{\Psi}$, listed in Table \ref{U(1)_X additional fields ele}, 
and the term $\tilde{\Phi}\Psi\tilde{\Psi}$ 
to the superpotential in the electric theory. The fields contribute 
to the 't Hooft anomalies by 
\be
&&\bullet~U(1)_XSU(N_f)^2:~0,
\qquad
\bullet~U(1)_X:~0,
\qquad
\bullet~U(1)_X^3:~-1+2\times\l(\hf\r)^3=-1+{1\over4},
\nn\\
&&\bullet~U(1)_XU(1)_R^2:~-1,
\qquad
\bullet~U(1)_XU(1)^2:~0,
\qquad
\bullet~U(1)_XU(1)U(1)_R:~0,
\nn\\
&&\bullet~U(1)_X^2U(1)_R:~-1,
\qquad
\bullet~U(1)_X^2U(1):~0,
\ne
which fill the gap in the anomalies $U(1)_X^3$, $U(1)_XU(1)_R^2$, 
$U(1)_X^2U(1)_R$. 

There still remain discrepancies in the anomalies 
$U(1)_XU(1)^2$ and $U(1)_XU(1)_RU(1)$. To fill the gap, 
we introduce two sets of fields $(\Xi, G, H)$, 
$(\tilde{\Xi}, \tilde{G}, \tilde{H})$, and add the terms 
$\Xi{GH}+\tilde{\Xi}\tilde{G}\tilde{H}$ to the superpotential 
in the electric theory. Their contributions to the 't Hooft anomalies 
are the following: 
\be
&&\bullet~U(1)_XSU(N_f)^2:~0,
\qquad
\bullet~U(1)_X:~0,
\qquad
\bullet~U(1)_X^3:~0,
\qquad
\bullet~U(1)_XU(1)_R^2:~0,
\nn\\
&&\bullet~U(1)_XU(1)^2:~2\times(-N_f)^2+2\times N_f^2=4N_f^2,
\nn\\
&&\bullet~U(1)_XU(1)_RU(1):~2\times1\times(-N_f)+2\times(-1)\times N_f=-4N_f,
\nn\\
&&\bullet~U(1)_X^2U(1)_R:~-8,
\qquad
\bullet~U(1)_X^2U(1):~0,
\ne
and we find that they fill the gap in the anomalies $U(1)_XU(1)^2$, 
$U(1)_XU(1)_RU(1)$. However, unfortunately, they make a gap again 
in the anomaly $U(1)_X^2U(1)_R$, and all the other anomalies 
satisfy the matching conditions. 

\begin{table}[htb]
\begin{center}
\begin{tabular}{|c|c||c|c|c||c|}\hline
 & $SU(N_f-5)$  & $SU(N_f)$ & $U(1)$ & $U(1)_R$  & $U(1)_X$
\\ \hline
$\tilde{q}_i$  & $\cpxfund$  & $\cpxfund$  
& $-2$ & ${8 \over N_f}-{1 / (N_f-5)}$ & $-1/(2N_f)-1/(2(N_f-5))$ 
\\ \hline
${q}$  & $\fund$ &   $\1$ 
& $2N_f$ &  $-1+{1 / (N_f-5)}$ & $1/(2(N_f-5))$
\\ \hline
$s$  & $\symm$ &     $\1$ 
& $0$ &  ${2 / (N_f-5)}$ & $1/(N_f-5)$
\\ \hline
$M^{ij}$  & ${\bf 1}$ &    $\symm$ & $4$ & $2-{16 / N_f}$ & $1/N_f$
\\ \hline
$Y^{i}$  & ${\bf 1}$ & $\fund$ & $-2(N_f-1)$ & $3-{8 / N_f}$ & $1/(2N_f)$
\\ \hline
$X$ & $\1$ & $\1$ & 0 & 0  & $-1$
\\ \hline\hline
$\tilde{X}'$  & $\1$  & $\1$  
& $0$ & $0$ & $-2$ 
\\ \hline
$F'$  & $\1$ &   $\1$ 
& $0$ &  $0$ & $2$
\\ \hline
$\tilde{F}'$  & $\1$ &     $\1$ 
& $0$ &  $2$ & $0$
\\ \hline
\end{tabular}
\end{center}
\caption{The original magnetic fields and 
the set of the fields added to the extended $SU(N_f-5)$ theory 
to fill the gaps in the continuous 't Hooft anomalies}
\label{U(1)_X additional fields mag}
\end{table}

In order to fill the gap in the anomaly $U(1)_X^2U(1)_R$, 
we will add three fields $\tilde{X}'$, $F'$, $\tilde{F}'$, 
listed in Table \ref{U(1)_X additional fields mag}, 
and add the term $\tilde{X}'F'\tilde{F}'$ to the superpotential 
in the magnetic extended theory. They contribute to the anomaly 
$U(1)_X^2U(1)_R$ by $-8$, and finally, all the anomalies 
with the $U(1)_X$ charges satisfy the continuous 't Hooft anomaly 
matching conditions.

However, the additional fields to the both sides cause 
the discrepancies in the anomalies $U(1)_R^3$ and 
$U(1)_R(\mbox{gravity})^2$. 
Upon the higgsing of the $U(1)_X$ symmetries by the vacuum expectation 
values $\vev{\Phi}$, $\langle{\tilde{\Phi}}\rangle$, $\vev{\Xi}$, 
$\langle{\tilde\Xi}\rangle$ 
on the electric side, $\vev{X}$, $\vev{X'}$ on the magnetic side, 
those extended theories are reduced into the original dual pair 
in the infrared, after the decoupling of the massive fields, 
and we find that the continuous 't Hooft anomalies without the $U(1)_X$ charges 
match with each other. Since the continuous 't Hooft anomalies with 
the $U(1)_X$ charges yields the discrete 't Hooft anomalies 
of the original dual theories in the infrared, it is obvious from 
the above computations that they satisfy the discrete 't Hooft anomaly 
matching conditions.

\newpage

\end{document}